\documentclass[a4paper,11pt,fleqn]{article}
\usepackage[utf8]{inputenc}
\pdfoutput=1

\usepackage{jheppub,float} 

\usepackage{bibentry}
\nobibliography*

\usepackage[normalem]{ulem}

\usepackage{xcolor}

\usepackage{framed}
\colorlet{shadecolor}{lightgray}

\usepackage{parskip}

\usepackage{tocloft}

\usepackage{hyperref}

\usepackage{chngcntr}
\counterwithout{equation}{section}

\usepackage{epigraph}

\newcommand{\p}{\partial}

\newcommand{\ttbar}{\text{T}\overline{\text{T}}}

\title{\boldmath Chiral Decoupling from Irrelevant Deformations}

\author[a]{Subhroneel Chakrabarti}
\author[b]{and Madhusudhan Raman}

\affiliation[a]{Institute of Mathematical Sciences, Homi Bhabha National Institute (HBNI)\\
	IV Cross Road, C.~I.~T.~Campus, Taramani, Chennai 600 113, Tamil Nadu, India}
\affiliation[b]{Department of Theoretical Physics, Tata Institute of Fundamental Research\\
	Homi Bhabha Road, Navy Nagar, Colaba, Mumbai 400 005, Maharashtra, India}

\emailAdd{subhroneelc@imsc.res.in}
\emailAdd{madhur@theory.tifr.res.in}

\abstract{We show that at the level of the equations of motion (i.e.~classically), in the limit in which the $ \ttbar $ deformation parameter is sent to infinity, the left- and right-chiral sectors of $ \ttbar $-deformed free theories decouple.}

\preprint{TIFR/TH/20-2}

\notoc 
\begin{document} 
	\maketitle
	\flushbottom
	
	\newpage 
	The last few years have witnessed a surge of interest in a class of soluble deformations of two-dimensional quantum field theories, among them the so-called $ \ttbar $ deformation \cite{Zamolodchikov:2004ce}. This irrelevant deformation is triggered by the composite operator
	\begin{equation}
	\mathrm{T} \overline{\mathrm{T}}=- \operatorname{det} T_{\mu \nu} \ .
	\end{equation}
	It is a curious feature of this deformation that a number of physically interesting quantities such as the deformed classical Lagrangian \cite{Cavaglia:2016oda,Bonelli:2018kik,Conti:2018jho}, the finite volume spectrum \cite{Smirnov:2016lqw}, the S-matrix \cite{Dubovsky:2012wk,Dubovsky:2013ira,Dubovsky:2017cnj,Rosenhaus:2019utc}, and the torus partition function \cite{Cardy:2018sdv,Datta:2018thy,Aharony:2018bad} are determined in terms of the data of the undeformed theory. $ \ttbar $ deformations have also been studied from the point of view of holography \cite{McGough:2016lol,Kraus:2018xrn,Taylor:2018xcy,Hartman:2018tkw,Caputa:2019pam}, little string theories \cite{Giveon:2017nie,Giveon:2017myj,Asrat:2017tzd,Giribet:2017imm,Chakraborty:2018kpr,Chakraborty:2018aji}, and in the context of supersymmetric quantum field theories \cite{Baggio:2018rpv,Chang:2018dge,Jiang:2019hux,Chang:2019kiu,Ferko:2019oyv,Coleman:2019dvf}. For a pedagogical introduction to $ \ttbar $ deformations, we invite the reader to consult \cite{Jiang:2019hxb} and references therein.
	
	In this short note, we focus on $ \ttbar $ deformations of free quantum field theories.
	
	\section*{Bosons}
	We start with a free boson in two dimensional Euclidean space.\footnote{We adopt the conventions of \cite{DiFrancesco:1997nk}.} It is described by the Lagrangian
	\begin{equation} \label{free}
	L_0 = \frac{1}{2} \p \phi \bar{\p} \phi \ .
	\end{equation}
	Consider $ \ttbar $ deformations of this Lagrangian, first discussed in \cite{Cavaglia:2016oda}. A well-known result in the literature shows that it is possible to resum these deformations to all orders in the $ \ttbar $ coupling constant $ \lambda $ to get the non-local Lagrangian
	\begin{equation}\label{eq:TTDeformedFreeBoson}
	L_{\lambda} = \frac{1}{\lambda} \left( -1+\sqrt{1+\lambda \, \p \phi \bar{\p} \phi } \right) \ .
	\end{equation}
	The above Lagrangian can be interpreted as describing the dynamics of a Nambu-Goto string in a three-dimensional target space 
	\begin{align}
	L_{\lambda}&=\frac{1}{\lambda} \sqrt{\operatorname{det}\left(\sum_{\mu=1}^{3} \partial_{\alpha} X^{\mu} \partial_{\beta} X^{\mu}\right)} \ ,
	\end{align}
	in the so-called static gauge, i.e.~when one makes the identifications
	\begin{equation}\label{key}
	X^{1} \rightarrow x^0 \ , \quad X^{2} \rightarrow x^1 \ , \quad X^{3} \rightarrow \sqrt{\lambda} \, \phi \ .
	\end{equation}
	Here, $ x^0 $ and $ x^1 $ are Euclidean worldsheet coordinates, while the boson $ \phi $ makes up the (single) transverse oscillator. 
	
	Given the tantalising square-root form of the above Lagrangian, it is natural to wonder if these deformations are in any way related to the physics of D-branes \cite{Chang:2018dge,Brennan:2019azg}. Indeed, the Lagrangian \eqref{eq:TTDeformedFreeBoson}, after an analytic continuation to Minkowski spacetime, can be written in a more suggestive form as
	\begin{equation}\label{eq:DiracBraneAction}
	L_{\mathrm{D}}=-\frac{1}{\lambda} \left(-1+\sqrt{-\operatorname{det}\left(\eta_{\mu \nu}+\lambda\,\partial_{\mu} \phi \partial_{\nu} \phi\right)}\right) \ .
	\end{equation}
	As is well-known, a $ p $-brane embedded in $ (D+1) $-dimensional Minkowski space will break translation invariance along directions orthogonal to it. The dynamics of Nambu-Goldstone modes corresponding to the spontaneously broken translation invariance are governed by the Dirac Lagrangian. For $ p = 1 $, i.e.~for a string, the Lagrangian \eqref{eq:DiracBraneAction} is the Dirac Lagrangian, with the tension $T$ of the string is related to $ \ttbar $ coupling $\lambda$ as
	\begin{equation} \label{tension}
	T \longleftrightarrow \frac{1}{\lambda} \ .
	\end{equation}
	To summarise, the $ \ttbar $-deformed free boson Lagrangian, upon analytic continuation, is just the Dirac Lagrangian for a string with tension inversely proportional to the $ \ttbar $ coupling $ \lambda $.\footnote{The relationship between Nambu-Goto strings and $ \ttbar $ deformations has also been studied in \cite{Beratto:2019bap}.}
	
	In this article we call to attention a curious feature of the above deformation in the infinite coupling limit (i.e.~$\lambda \rightarrow \infty$) of a $ \ttbar $ deformation. In particular, we consider the $\lambda \rightarrow \infty$ limit \emph{after} the $ \ttbar $-deformed free boson Lagrangian \eqref{eq:TTDeformedFreeBoson} has been analytically continued to Minkowski spacetime. To the best of our knowledge, such a limit has remained unexplored in the literature. The tensionless limit of the Dirac Lagrangian, however, has been recently studied by \cite{Townsend:2019koy}, and we draw on some of the results of these investigations, bringing them to bear upon the study of $ \ttbar $ deformations via the aforementioned correspondence.
	
	The Hamiltonian corresponding to the Dirac Lagrangian is
	\begin{equation}
	\label{eq:DiracHamiltonian}
	H_{\text{D}}=\frac{1}{\lambda} \left( -1+\sqrt{1+\lambda\left[\pi^{2}+\left(\phi^{\prime}\right)^{2}\right]+\lambda^{2}\left[\phi^{\prime} \pi\right]^{2}} \right) \ ,
	\end{equation}
	where $ \pi $ is the canonical momentum conjugate to $ \phi $.\footnote{While this note was being completed, we came across \cite{Jorjadze:2020ili}, where $ \ttbar $ deformations are also considered within the Hamiltonian framework.} In the above equations, the dots are time derivatives and the primes are spatial derivatives, as usual. While the zero-coupling limit of \eqref{eq:DiracHamiltonian} returns the Hamiltonian for a free boson, the infinite-coupling limit yields
	\begin{equation}
	\lim _{\lambda \rightarrow \infty} H_{\text{D}}=\left|\phi^{\prime} \pi\right| \ ,
	\end{equation}
	and so the equations of motion in the infinite-coupling limit depend on the sign of $ \pi\phi' $. We argue that both choices of sign correspond to co-existent sectors of the theory. The equations of motion for each of these sector can be written down provided we impose the constraint $\pi = \pm \phi'$, only to be imposed at the level of equations of motion. It is easy, then, to verify using the equations of motion that this choice of sign $ \pi=\pm \phi' $ implies in turn a choice of chirality. 
	
	We conclude from this that the classical Hamiltonian breaks up into two sectors describing left-chiral and right-chiral bosons. In other words, the infinite-coupling limit effects a decoupling of the chiral halves of the boson. It is impossible to simultaneously write down the equations of motion for both sectors; nontheless, the two sectors are straightforwardly related by a parity transformation. 
	
	Instead of describing the chiral bosons using their equations of motions, we may wish to make the chirality of a specific sector manifest at the level of the Hamiltonian. For example, we find that with (say) the choice $ \pi=+ \phi^{\prime} $, we have
	\begin{equation}
	H_{\text{D}} = \left(\phi'\right)^2 \ ,
	\end{equation}
	and the resulting Lagrangian in configuration space is
	\begin{equation}\label{FJ}
	L_{\text{FJ}} = \phi' \left(\dot{\phi} - \phi'\right) \ .
	\end{equation}
	The Lagrangian \eqref{FJ} is the well-known Floreanini-Jackiw Lagrangian for a chiral boson \cite{Floreanini:1987as}. Therefore, we conclude that at the level of classical Lagrangians, the infinite coupling limit of a $ \ttbar $-deformed free boson effects a decoupling of the left-chiral and right-chiral halves of the bosons. 
	
	At this stage, it may be tempting to conclude that the infinite coupling limit halves the number of degrees of freedom and leaves us with a chiral boson. If this were true, only one of the two possible signs of $ \pi \phi' $ is physical, not both. This conclusion is incorrect. The infinite coupling limit effectively maps the original (free) theory to a theory in which both chiralities which are decoupled. There is no reason to privilege one choice of the sign over the other. Additionally, there is nothing to suggest that on taking the infinite coupling limit, a parity-invariant theory should go into a parity-violating theory.  Finally, this conclusion is consistent with expectations from considerations pertaining to the gravitational anomaly. The theory \eqref{eq:TTDeformedFreeBoson}, when coupled to a metric, is anomaly free. If and only if the infinite coupling limit returns both left- and right-chiral sectors will the final theory, on coupling to a metric, be free of gravitational anomalies.\footnote{We couple to the metric only \textit{after} performing $ \ttbar $ deformations and taking the infinite coupling limit.} Finally, since the infinite-coupling limit does not seem to affect any considerations pertaining to gravitational anomaly, it is natural to expect that both the initial and the final theories, on coupling to a metric, yield a theory devoid of gravitational anomalies.
	
	It is well known that equal numbers of left- and right-chiral bosons cannot always be thought of as describing ordinary bosons. The observation central to this note provides a novel (but not unique) example contrary to this common expectation. This is guaranteed because the $ \ttbar $ deformation is an invertible deformation. In principle, given a pair of left- and right-chiral bosons, we can always reverse the steps taken in this note, i.e.~restore the coupling and run the $ \ttbar $ deformation machinery in reverse to obtain a theory of ordinary free bosons. Other instances in which a scalar is known to undergo chiral decoupling are when the scalar is either compact or twisted. It seems that the chiral decoupling presented in this article is unrelated to either of those known instances of chiral decoupling.

	Often in the literature the $ \ttbar $ deformation is thought of as a ``reversal of the renormalization group flow." This is true in the sense that the theory \eqref{eq:TTDeformedFreeBoson}, when thought of as an UV theory, under the renormalization group flow ends up in an IR fixed point given by the free boson \eqref{free}. If one thinks of the infinite-coupling limit as a part of the flow, albeit towards the UV starting from \eqref{eq:DiracBraneAction}, then one observes the following curious feature: a local theory of chirally decoupled bosons first flows down to a \emph{non-local} interacting theory of a regular boson, and then flows down to a local theory of regular boson interacting via derivative interactions, to eventually a local theory of free bosons in the IR.\footnote{We acknowledge helpful correspondence with Shouvik Datta regarding this point.}
	
	\section*{Fermions}	
	Analogously, one may study the $ \ttbar $-deformed free fermion in the limit of infinite coupling. We start out in Euclidean space, as before. The $ \ttbar $-deformed Lagrangian for a free fermion is\footnote{This Lagrangian has been shown in \cite{Cribiori:2019xzp} to be equivalent to the $ N = (2,2) $ Volkov-Akulov Lagrangian.}
	\begin{equation}
	L_\lambda = \psi_-^\dagger \p \psi_- + \psi_+^\dagger \bar{\p} \psi_+ - \lambda \left[\left(\psi_-^\dagger \p \psi_-\right)\left(\psi_+^\dagger \bar{\p} \psi_+\right) - \left(\psi_-^\dagger \bar{\p} \psi_-\right)\left(\psi_+^\dagger \p \psi_+\right)\right] \ . \label{fermion_deformed}
	\end{equation}
	After analytic continuation, we compute the Hamiltonian in the usual way, via a Legendre transform. The result is
	\begin{equation}
	H_{\lambda} = -\frac{1}{2\lambda} \left(-1 + \sqrt{1+4\lambda \left[\Pi_- \psi_-' - \Pi_+ \psi_+' \right]+4\lambda^2 \left[\Pi_- \psi_-' + \Pi_+ \psi_+'\right]^2}\right) \ ,
	\end{equation}
	where primes denote spatial derivatives, and $ \Pi_{\pm} $ are the canonical momenta conjugate to $ \psi_{\pm} $. In the limit of vanishing $ \lambda $, this returns 
	\begin{equation}
	H_0 = \lim_{\lambda\rightarrow 0} H_\lambda =  - \Pi_- \psi_-' + \Pi_+ \psi_+' \ ,
	\end{equation}
	as one would expect. At large $ \lambda $, however, one gets
	\begin{equation}
	H_{\infty} = \lim_{\lambda\rightarrow \infty} H_\lambda = -\left|\Pi_- \psi_-' + \Pi_+ \psi_+'\right| \ ,
	\end{equation}
	and the corresponding first-order action is
	\begin{equation}
	S = \int \text{d}^2 \sigma \left[ \Pi_+ \dot{\psi}_+ + \Pi_- \dot{\psi}_- -   H_{\infty}    \right] \ .
	\end{equation}
	At this stage itself, we can conclude that the infinite-coupling limit yields decoupled chiral fermions. Let us be a little more explicit, for clarity.
	
	In a form structurally reminiscent of the bosons, the equations of motion in the infinite-coupling limit depend on a sign, in this case of $\left(\Pi_- \psi_-' + \Pi_+ \psi_+'\right)$ . There are, once again, two sectors corresponding to two possible choices of sign, and they are described by the following actions: for the $ (+) $ sign, we have
	\begin{align}
	H_\infty &= \left(\Pi_- \psi_-' + \Pi_+ \psi_+'\right) \ , \nonumber \\
	S_{\text{R}} &= \int \text{d}^2 \sigma \left[ \Pi_+ \partial_- \psi_+ + \Pi_- \partial_- \psi_-  \right] \ , \label{sright}
	\end{align}
	and for the $ (-) $ sign, we have
	\begin{align}
	H_\infty &= -\left(\Pi_- \psi_-' + \Pi_+ \psi_+'\right) \,, \nonumber \\
	S_{\text{L}} &= \int \text{d}^2 \sigma \left[ \Pi_+ \partial_+ \psi_+ + \Pi_- \partial_+ \psi_-  \right] \, . \label{sleft}
	\end{align}
	From the actions \eqref{sright} and \eqref{sleft} we conclude that in either sector, the two chiralities are decoupled. 
	
	In close analogy with the bosons, we may choose to impose constraints on the equations of motions that resolve the sign ambiguity introduced by the absolute value. A suitable pair of constraints is 
	\begin{align}
	\textsc{Right Sector:} \quad \Pi_- &= 0 \quad \text{and} \quad \Pi_+ \psi_+^{'} > 0 \,\,, \label{right} \\
	\textsc{Left Sector:} \quad \Pi_+ &= 0 \quad \text{and} \quad \Pi_- \psi_-^{'} > 0 \,\,. \label{left}
	\end{align}
	The choice of setting the non-zero term in $H_\infty$ to be \emph{less} than zero forces us to conclude that $\psi_-$ propagates in the $(+)$ direction, and likewise for $ \psi_+ $ and the $ (-) $ direction. We reject these branches on the basis of a definite convention: the chiral fermion $\psi_{+}$ (respectively, $ \psi_- $) propagates along the $(+)$ (respectively, $ (-) $) direction. 
	
	It is once again straightforward to see that the constraints \eqref{right} and \eqref{left} imposed on the equations of motion obtained from the actions \eqref{sright} and \eqref{sleft} respectively are nothing but chirality constraints, describing the classical dynamics of chiral fermions. Once again the two sectors are simply related to each other by a parity transformation.
	
	\section*{Further Comments}
	
	It is interesting to compare and contrast our findings with what is already known about the $ \ttbar $ deformation. One of the beautiful aspects of these deformations is the ability to determine the exact finite-volume spectrum of the deformed theory. Consider a quantum field theory wrapped around a cylinder of circumference $ R $, whose undeformed energies and quantised momenta are $ E_n $ and $ P_n $ respectively. The energy spectrum is deformed by the $ \ttbar $ deformation, while the momenta remain unchanged. The deformed spectrum $ E^{\lambda}_n $ is
	\begin{equation}
	\label{eq:DeformedSpectrum}
	E^{\lambda}_n = \frac{R}{\lambda} \left(-1+\sqrt{1+\frac{\lambda\,E_n}{R} + \frac{\lambda^2 \, P_n^2}{R^2}} \right) \ .
	\end{equation}
	On considering the above result in the limit $ \lambda \rightarrow \infty $, we find
	\begin{equation}
	E^{\infty}_n = \left|P_n\right| \ ,
	\end{equation}
	which is the dispersion relation of a classical \emph{chiral} field. This moves us to conclude that the chiral decoupling we have discussed in this paper may be generic---in this regard, free field theories are only the simplest and most transparent examples of this decoupling. Further, since the deformed spectrum is quantum mechanically exact, it is likely that chiral decoupling at infinite $ \lambda $ is a generic feature of $ \ttbar $-deformed quantum field theories.
	
	Another attractive feature of these irrelevant deformations are that the exact S-matrix of the deformed theory is arrived at by multiplying the S-matrix of the undeformed theory by a phase. For the scattering of massless particles $ (i,j) \rightarrow (k,l) $ with relative rapidity $ \theta = \theta_i - \theta_j $, the S-matrix changes as
	\begin{equation}
	S_{ij}^{kl}(\theta) \overset{\ttbar}{\longrightarrow} S_{ij}^{kl}(\theta) \, e^{i \delta_{i j}^{(\lambda)}(\theta)} \ ,
	\end{equation}
	where the phase is referred to as the CDD factor, and takes the form
	\begin{equation}
	\delta_{i j}^{(\lambda)}\left(\theta\right) = -2 \lambda \, p_{i}^{(+)} p_{j}^{(-)} \ ,
	\end{equation}
	where $ p_{i}^{(+)} $ and $ p_{j}^{(-)} $ are the momenta of the left- and right-moving particles. Implicit in the derivation of the CDD factor is the assumption that $ \lambda $ is finite. Since the coupling $ \lambda $ only appears in the phase, rigorously taking the limit $ \lambda \rightarrow \infty $ (with other parameters held fixed) of the CDD factor is at best ambiguous. One possible resolution to this ambiguity would be to start with the classical theory at infinite coupling, following the analysis presented in this paper, and then compute the S-matrix. It would be interesting to better understand how the infinite coupling limit should be interpreted in the context of scattering.
	
	Finally, irrespective of the imposition of the constraints, the chiral decoupling of fermions in the infinite-coupling limit represents a curious development. The original action \eqref{fermion_deformed} appears to be an interacting theory with coupling constant $\lambda$. In the limit where the coupling constant goes to infinity --- when the interaction term is dominant --- we find that the theory breaks up into two sectors, each describing a chiral fermion. 
	
	Indeed, the same is true of the bosonic case as well: the theory is interacting at finite, non-zero values of coupling and when the coupling constant goes to infinity, the theory breaks up into two sectors, each describing a chiral boson. 
	
	How might one explain this development? Since our analyses in this note are purely based on equations of motion, it would be interesting to study the corresponding quantum field theories, perhaps along the lines of \cite{Rosenhaus:2019utc,Callebaut:2019omt}. We hope to return to these questions in the near future.

	\acknowledgments{We are grateful to Sujay Ashok, Divyanshu Gupta, and Arkajyoti Manna for collaboration and discussions on a related project, to Shouvik Datta and Ashoke Sen for helpful correspondence, and especially to Dileep Jatkar for encouragement, suggestions, and a careful reading of an earlier version of this preprint. We are also thankful to the referee for their insightful comments. MR acknowledges support from the Infosys Endowment for Research into the Quantum Structure of Spacetime.}
	
	\bibliographystyle{JHEP}
	\bibliography{Refs}

\providecommand{\href}[2]{#2}\begingroup\raggedright\begin{thebibliography}{10}

\bibitem{Zamolodchikov:2004ce}
A.~B. Zamolodchikov, {\it {Expectation value of composite field T anti-T in
  two-dimensional quantum field theory}},
  \href{http://arxiv.org/abs/hep-th/0401146}{{\tt hep-th/0401146}}.

\bibitem{Cavaglia:2016oda}
A.~Cavaglià, S.~Negro, I.~M. Szécsényi, and R.~Tateo, {\it {$T
  \bar{T}$-deformed 2D Quantum Field Theories}},  {\em JHEP} {\bf 10} (2016)
  112, [\href{http://arxiv.org/abs/1608.05534}{{\tt arXiv:1608.05534}}].

\bibitem{Bonelli:2018kik}
G.~Bonelli, N.~Doroud, and M.~Zhu, {\it {$T \bar{T}$-deformations in closed
  form}},  {\em JHEP} {\bf 06} (2018) 149,
  [\href{http://arxiv.org/abs/1804.10967}{{\tt arXiv:1804.10967}}].

\bibitem{Conti:2018jho}
R.~Conti, L.~Iannella, S.~Negro, and R.~Tateo, {\it {Generalised Born-Infeld
  models, Lax operators and the $ \mathrm{T}\overline{\mathrm{T}} $
  perturbation}},  {\em JHEP} {\bf 11} (2018) 007,
  [\href{http://arxiv.org/abs/1806.11515}{{\tt arXiv:1806.11515}}].

\bibitem{Smirnov:2016lqw}
F.~A. Smirnov and A.~B. Zamolodchikov, {\it {On space of integrable quantum
  field theories}},  {\em Nucl. Phys.} {\bf B915} (2017) 363--383,
  [\href{http://arxiv.org/abs/1608.05499}{{\tt arXiv:1608.05499}}].

\bibitem{Dubovsky:2012wk}
S.~Dubovsky, R.~Flauger, and V.~Gorbenko, {\it {Solving the Simplest Theory of
  Quantum Gravity}},  {\em JHEP} {\bf 09} (2012) 133,
  [\href{http://arxiv.org/abs/1205.6805}{{\tt arXiv:1205.6805}}].

\bibitem{Dubovsky:2013ira}
S.~Dubovsky, V.~Gorbenko, and M.~Mirbabayi, {\it {Natural Tuning: Towards A
  Proof of Concept}},  {\em JHEP} {\bf 09} (2013) 045,
  [\href{http://arxiv.org/abs/1305.6939}{{\tt arXiv:1305.6939}}].

\bibitem{Dubovsky:2017cnj}
S.~Dubovsky, V.~Gorbenko, and M.~Mirbabayi, {\it {Asymptotic fragility, near
  AdS$_{2}$ holography and $ T\overline{T} $}},  {\em JHEP} {\bf 09} (2017)
  136, [\href{http://arxiv.org/abs/1706.06604}{{\tt arXiv:1706.06604}}].

\bibitem{Rosenhaus:2019utc}
V.~Rosenhaus and M.~Smolkin, {\it {Integrability and Renormalization under $T
  \bar T$}},  \href{http://arxiv.org/abs/1909.02640}{{\tt arXiv:1909.02640}}.

\bibitem{Cardy:2018sdv}
J.~Cardy, {\it {The $ T\overline{T} $ deformation of quantum field theory as
  random geometry}},  {\em JHEP} {\bf 10} (2018) 186,
  [\href{http://arxiv.org/abs/1801.06895}{{\tt arXiv:1801.06895}}].

\bibitem{Datta:2018thy}
S.~Datta and Y.~Jiang, {\it {$T\bar{T}$ deformed partition functions}},  {\em
  JHEP} {\bf 08} (2018) 106, [\href{http://arxiv.org/abs/1806.07426}{{\tt
  arXiv:1806.07426}}].

\bibitem{Aharony:2018bad}
O.~Aharony, S.~Datta, A.~Giveon, Y.~Jiang, and D.~Kutasov, {\it {Modular
  invariance and uniqueness of $T\bar{T}$ deformed CFT}},  {\em JHEP} {\bf 01}
  (2019) 086, [\href{http://arxiv.org/abs/1808.02492}{{\tt arXiv:1808.02492}}].

\bibitem{McGough:2016lol}
L.~McGough, M.~Mezei, and H.~Verlinde, {\it {Moving the CFT into the bulk with
  $ T\overline{T} $}},  {\em JHEP} {\bf 04} (2018) 010,
  [\href{http://arxiv.org/abs/1611.03470}{{\tt arXiv:1611.03470}}].

\bibitem{Kraus:2018xrn}
P.~Kraus, J.~Liu, and D.~Marolf, {\it {Cutoff AdS$_{3}$ versus the $
  T\overline{T} $ deformation}},  {\em JHEP} {\bf 07} (2018) 027,
  [\href{http://arxiv.org/abs/1801.02714}{{\tt arXiv:1801.02714}}].

\bibitem{Taylor:2018xcy}
M.~Taylor, {\it {TT deformations in general dimensions}},
  \href{http://arxiv.org/abs/1805.10287}{{\tt arXiv:1805.10287}}.

\bibitem{Hartman:2018tkw}
T.~Hartman, J.~Kruthoff, E.~Shaghoulian, and A.~Tajdini, {\it {Holography at
  finite cutoff with a $T^2$ deformation}},  {\em JHEP} {\bf 03} (2019) 004,
  [\href{http://arxiv.org/abs/1807.11401}{{\tt arXiv:1807.11401}}].

\bibitem{Caputa:2019pam}
P.~Caputa, S.~Datta, and V.~Shyam, {\it {Sphere partition functions \& cut-off
  AdS}},  {\em JHEP} {\bf 05} (2019) 112,
  [\href{http://arxiv.org/abs/1902.10893}{{\tt arXiv:1902.10893}}].

\bibitem{Giveon:2017nie}
A.~Giveon, N.~Itzhaki, and D.~Kutasov, {\it {$ \mathrm{T}\overline{\mathrm{T}}
  $ and LST}},  {\em JHEP} {\bf 07} (2017) 122,
  [\href{http://arxiv.org/abs/1701.05576}{{\tt arXiv:1701.05576}}].

\bibitem{Giveon:2017myj}
A.~Giveon, N.~Itzhaki, and D.~Kutasov, {\it {A solvable irrelevant deformation
  of AdS$_{3}$/CFT$_{2}$}},  {\em JHEP} {\bf 12} (2017) 155,
  [\href{http://arxiv.org/abs/1707.05800}{{\tt arXiv:1707.05800}}].

\bibitem{Asrat:2017tzd}
M.~Asrat, A.~Giveon, N.~Itzhaki, and D.~Kutasov, {\it {Holography Beyond AdS}},
   {\em Nucl. Phys.} {\bf B932} (2018) 241--253,
  [\href{http://arxiv.org/abs/1711.02690}{{\tt arXiv:1711.02690}}].

\bibitem{Giribet:2017imm}
G.~Giribet, {\it {$T\bar{T}$-deformations, AdS/CFT and correlation functions}},
   {\em JHEP} {\bf 02} (2018) 114, [\href{http://arxiv.org/abs/1711.02716}{{\tt
  arXiv:1711.02716}}].

\bibitem{Chakraborty:2018kpr}
S.~Chakraborty, A.~Giveon, N.~Itzhaki, and D.~Kutasov, {\it {Entanglement
  beyond AdS}},  {\em Nucl. Phys.} {\bf B935} (2018) 290--309,
  [\href{http://arxiv.org/abs/1805.06286}{{\tt arXiv:1805.06286}}].

\bibitem{Chakraborty:2018aji}
S.~Chakraborty, {\it {Wilson loop in a $T\bar{T}$ like deformed $\rm{CFT}_2$}},
   {\em Nucl. Phys.} {\bf B938} (2019) 605--620,
  [\href{http://arxiv.org/abs/1809.01915}{{\tt arXiv:1809.01915}}].

\bibitem{Baggio:2018rpv}
M.~Baggio, A.~Sfondrini, G.~Tartaglino-Mazzucchelli, and H.~Walsh, {\it {On $
  T\overline{T} $ deformations and supersymmetry}},  {\em JHEP} {\bf 06} (2019)
  063, [\href{http://arxiv.org/abs/1811.00533}{{\tt arXiv:1811.00533}}].

\bibitem{Chang:2018dge}
C.-K. Chang, C.~Ferko, and S.~Sethi, {\it {Supersymmetry and $ T\overline{T} $
  deformations}},  {\em JHEP} {\bf 04} (2019) 131,
  [\href{http://arxiv.org/abs/1811.01895}{{\tt arXiv:1811.01895}}].

\bibitem{Jiang:2019hux}
H.~Jiang, A.~Sfondrini, and G.~Tartaglino-Mazzucchelli, {\it {$T\bar{T}$
  deformations with $\mathcal{N}=(0,2)$ supersymmetry}},  {\em Phys. Rev.} {\bf
  D100} (2019), no.~4 046017, [\href{http://arxiv.org/abs/1904.04760}{{\tt
  arXiv:1904.04760}}].

\bibitem{Chang:2019kiu}
C.-K. Chang, C.~Ferko, S.~Sethi, A.~Sfondrini, and G.~Tartaglino-Mazzucchelli,
  {\it {$T\bar{T}$ Flows and (2,2) Supersymmetry}},
  \href{http://arxiv.org/abs/1906.00467}{{\tt arXiv:1906.00467}}.

\bibitem{Ferko:2019oyv}
C.~Ferko, H.~Jiang, S.~Sethi, and G.~Tartaglino-Mazzucchelli, {\it {Non-Linear
  Supersymmetry and $T\bar T$-like Flows}},
  \href{http://arxiv.org/abs/1910.01599}{{\tt arXiv:1910.01599}}.

\bibitem{Coleman:2019dvf}
E.~A. Coleman, J.~Aguilera-Damia, D.~Z. Freedman, and R.~M. Soni, {\it {$
  T\overline{T} $ -deformed actions and (1,1) supersymmetry}},  {\em JHEP} {\bf
  10} (2019) 080, [\href{http://arxiv.org/abs/1906.05439}{{\tt
  arXiv:1906.05439}}].

\bibitem{Jiang:2019hxb}
Y.~Jiang, {\it {Lectures on solvable irrelevant deformations of 2d quantum
  field theory}},  \href{http://arxiv.org/abs/1904.13376}{{\tt
  arXiv:1904.13376}}.

\bibitem{DiFrancesco:1997nk}
P.~Di~Francesco, P.~Mathieu, and D.~Senechal, {\em {Conformal Field Theory}}.
\newblock Graduate Texts in Contemporary Physics. Springer-Verlag, New York,
  1997.

\bibitem{Brennan:2019azg}
T.~D. Brennan, C.~Ferko, and S.~Sethi, {\it {A Non-Abelian Analogue of DBI from
  $T \overline{T}$}},  \href{http://arxiv.org/abs/1912.12389}{{\tt
  arXiv:1912.12389}}.

\bibitem{Beratto:2019bap}
E.~Beratto, M.~Billo', and M.~Caselle, {\it {On the $T\bar T$ deformation of
  the compactified boson and its interpretation in Lattice Gauge Theory}},
  \href{http://arxiv.org/abs/1912.08654}{{\tt arXiv:1912.08654}}.

\bibitem{Townsend:2019koy}
P.~K. Townsend, {\it {A manifestly Lorentz invariant chiral boson action}},
  \href{http://arxiv.org/abs/1912.04773}{{\tt arXiv:1912.04773}}.

\bibitem{Jorjadze:2020ili}
G.~Jorjadze and S.~Theisen, {\it {Canonical maps and integrability in $T\bar T$
  deformed 2d CFTs}},  \href{http://arxiv.org/abs/2001.03563}{{\tt
  arXiv:2001.03563}}.

\bibitem{Floreanini:1987as}
R.~Floreanini and R.~Jackiw, {\it {Selfdual Fields as Charge Density
  Solitons}},  {\em Phys. Rev. Lett.} {\bf 59} (1987) 1873.

\bibitem{Cribiori:2019xzp}
N.~Cribiori, F.~Farakos, and R.~von Unge, {\it {The 2D Volkov-Akulov model as a
  $T \bar{T}$ deformation}},  {\em Phys. Rev. Lett.} {\bf 123} (2019), no.~20
  201601, [\href{http://arxiv.org/abs/1907.08150}{{\tt arXiv:1907.08150}}].

\bibitem{Callebaut:2019omt}
N.~Callebaut, J.~Kruthoff, and H.~Verlinde, {\it {$T\bar{T}$ deformed CFT as a
  non-critical string}},  \href{http://arxiv.org/abs/1910.13578}{{\tt
  arXiv:1910.13578}}.

\end{thebibliography}\endgroup
\end{document}